\documentclass[10pt,conference]{IEEEtran}
\IEEEoverridecommandlockouts
\usepackage{cite}
\usepackage{amsmath,amssymb,amsfonts}
\usepackage{algorithmic}
\usepackage{graphicx}
\usepackage{textcomp}
\usepackage{xcolor}
\usepackage{caption}
\usepackage{textgreek}
\usepackage{subcaption}
\def\BibTeX{{\rm B\kern-.05em{\sc i\kern-.025em b}\kern-.08em
    T\kern-.1667em\lower.7ex\hbox{E}\kern-.125emX}}
\begin{document}

\title{Binary Code Similarity Detection
}

\author{\IEEEauthorblockN{Zian Liu}
\IEEEauthorblockA{\textit{Department of Computer Science and Software Engineering} \\
\textit{Swinburne University of Technology}\\
\textit{Data61, CSIRO}\\
Melbourne, Australia \\
102516622@student.swin.edu.au}
}

\maketitle



\begin{abstract}
Binary code similarity detection is to detect the similarity of code at binary (assembly) level without source code. Existing works have their limitations when dealing with mutated binary code generated by different compiling options. 
In this paper, we propose a novel approach to addressing this problem. By inspecting the binary code, we found that generally, within a function, some instructions aim to calculate (prepare) values for other instructions. The latter instructions are defined by us as \emph{key} instructions. Currently, we define four categories of key instructions: calling subfunctions, comparing instruction, returning instruction, and memory-store instruction. Thus if we symbolically execute similar binary codes, symbolic values at these key instructions are expected to be similar. As such, we implement a prototype tool, which has three steps. \emph{First}, it symbolically executes  binary code; \emph{Second}, it extracts symbolic values at defined key instructions into a graph; \emph{Last}, it compares the symbolic graph similarity. In our implementation, we also address some problems, including path explosion and loop handling.  
\end{abstract}


\section{Introduction}
My PhD goal aims at analyzing binary code to improve binary code security. This includes similarity detection, code diffing, bug finding, etc., at binary level. I plan to achieve this goal by implementing novel approaches in a binary code similarity detecting tool. The current process is that the tool's framework has finished. I am fine-tuning this tool to conducting on a large scale dataset.

With the rapid development of the software industry, binary code similarity detection is playing an increasingly critical role. Software producers release their products as binary code mainly to protect their source code. Binary code similarity detection can be used in many fields such as bug search, malware detection, malware clustering, malware linage, patch generation and analysis, porting information, software theft detection\cite{binarysurvey}, etc. However, binary code similarity detection is challenging, which is mainly due to, 1) The same source code can be compiled on different architectures, resulting in different binary codes using different instruction sets; 2) During compiling process, different compilers and compiling options may produce significantly different binary code for the same source code; 3) Producers can use many code protecting tools to obfuscate the code. There are some existing solutions to this problem. These works can be divided into three categories with different standards. In Section \ref{background}, we mainly introduce these works according to their comparison type: syntactic, semantic, and structural similarity. A typical syntactical approach represents the code by statisticizing the occurrence of specific strings in the code. Alternatively, one can use machine learning based method to learn the vectorized representation of one or more sentences in the code. A semantic approach evaluates whether two codes have similar functionality or impact. Structural approach compares graph features such as control flow graph and call graph. However, these solutions still have some limitations. For example, syntactic similarity focus on the `appearance' of binary code thus sometimes failing to explore semantic similarity. Semantic similarity, however, is not scalable and may not cover all of the code or situations. Moreover, Ren et al. pointed out that existing binary code similarity detection overlooked the impact of compiling options \cite{unleashing}. They claimed that the testing set used in current work lacked enough compiling options mutation. They tested several state-of-the-art binary similarity detection tools with their crafted dataset. The detecting result all turned out to be less accurate. Structural similarity is vulnerable for cross-architecture comparison since they assume that basic blocks remain their feature and relationship with other basic blocks under all circumstances. However, by utilizing Ren el al's method, one can tune the compiling configuration to have less structural similarity. Therefore, our research question is how can we detect binary code similarity even when these compiling options are deployed.

By analyzing the binary code, we observed that some instructions calculate value for some other instructions. The latter instructions are more important for deciding the similarity while the former instructions should be given less consideration. This is because the symbolic value of the former can propagate to latter instructions. Only inspecting the symbolic value of these latter instructions can reserve the main symbolic meaning of the function. We thus define these important instructions as Key Instructions. In our tool, they are translated to Intermediate Representation (IR) and connected based on control flow to form a Key IR graph. We implemented our idea in a prototype tool. This tool is able to lift different architectures' binary code into Key IR graphs and compare their similarity. This tool consists of three phases: 1) symbolic execution, 2) Key IR graph construction, and 3) Key IR graph similarity detection. In summary, our contributions are: 1) We proposed a Key IR graph to abstract the semantic of binary code. 2) We implemented a prototype tool, including the symbolic execution engine, IR lifting engine, and comparison engine. This tool currently supports two architectures, i.e., x86-64 and ARM. 3) We proposed a novel method to compare the similarity of Key IR graphs. Currently we have finished the implementation process and on our way to evaluation.

\section{Background}
\label{background}
As \cite{binarysurvey} mentioned in their survey, existing works on binary code similarity can be classified according to different standards. In terms of the comparison granularity, these works can be divided into instruction level, basic block level, function level, and whole program level. In the binary similarity comparison, one input is used as a query to compare with the target. Thus according to input-target number, they can be classified into one-to-one, one-to-many, and many-to-many comparisons. According to supported architectures, they can be divided into a single architecture and cross-architecture. Single architecture only supports one kind of assembly instruction set, while cross-architecture supports more than one. According to analysis type, they can be categorized into static analysis, static analysis, and hybrid analysis. Static analysis analyzes the code without running the code, while dynamic analysis executes the code with some input. The advantage of static analysis is scalability and code coverage. Dynamic analysis is resource consuming but helpful to gain semantic information. According to comparison type, they can be classified into syntax, structural, and semantics similarity comparison. Syntax similarity captures instruction representation similarity, while structural similarity focuses on similarity in terms of graph representation of the binary code. Semantic similarity highlights the similarity of code's impact. 
 
We introduce various techniques used according to different comparison types. For syntax similarity, common strategies include hashing, embedding, and alignment. \cite{SMIT,BINCLONE,SPAIN} all use hashing technique to output various instructions sequences into fixed length of output hashing value. Equivalent output value implies syntax similarity. \cite{IDEA,mbc,Expose,Kam1n0} generate an embedding from sequences. \cite{aDiff,innereye,asm2vec,safe} automatically learn the embeddings for each instruction and use them to produce the basic-level or function-level embedding. \cite{exediff,tracy,binsequence} align two sequences and decide their similarity. For structural similarity, common methods are optimization solution, k-subgraph matching, path similarity, and graph embedding. \cite{SMIT,binslayer,cxz2014,genius} transform the problem into finding the mapping between two CFGs with minimum cost (optimization solution). \cite{BEAGLE,cxz2014,rendezvous,fossil} divide the graph into k subgraphs, so that each one has k connected nodes. Matched number of those subgraphs indicates the extent of the similarity. \cite{COP,SIGMA,binsequence} determine similarity based on paths. \cite{genius,VULSEEKER,GIMINI} extract features from graphs into feature vectors and determine the vector similarity. For semantic similarity, general techniques are instruction classification, input-output pairs, symbolic execution, theorem prover, and semantic hashes. \cite{BEAGLE,fossil,SIGMA} classified instruction based on their semantic in terms of arithmetic,
logic or data transfer purpose. \cite{binhash,multimh,bingo,SPAIN,ks2017,IMF-SIM} check whether output are the same to the input. \cite{binhunt,binhash,Expose,COP,multimh,esh} use symbolic formulas to represent the binary code. With the symbolic formula, \cite{binhunt} uses Theorem Prover to check whether two different symbolic formula has similar output. \cite{binhash,binjuice,GITZ} use symbolic hashes as an alternative to the theorem prover. \cite{xmatch,tedem} determine the edit distance of the tree/graph of the symbolic formula.

\section{Methodology}
The comparison granularity of our tool is function level. Given two functions from two different input binary codes, our tool consists of three phases: 1. symbolic execution, 2. Key IR graph construction, and 3. Key IR graph similarity comparison. We firstly symbolically execute the binary code several times to get their possible symbolic values. Then for the most important instructions for comparison, we lift them and their symbolic values into Key IRs. Those Key IRs are then combined to produce the Key IR graph. Lastly, we compare the similarity of two IR graphs.

\subsection{Symbolic execution}

\begin{figure}
\centerline{\includegraphics[width=20mm]{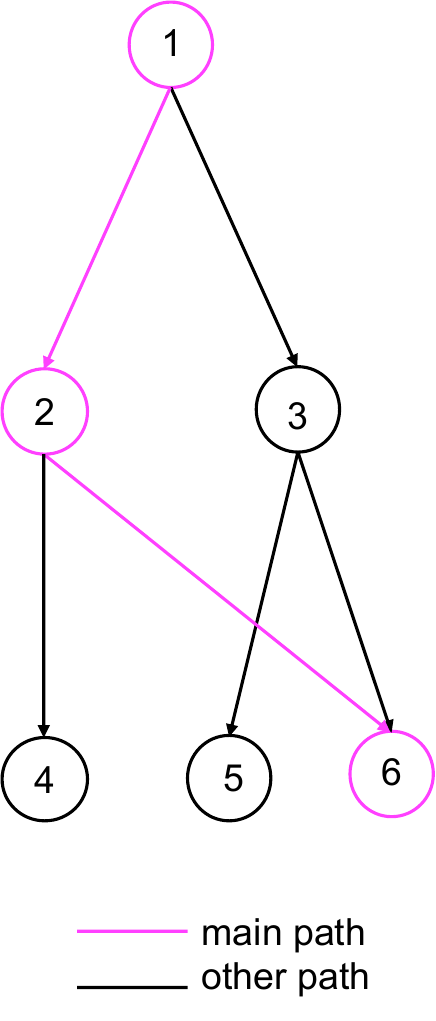}}
\caption{1st run}
\label{main-other-path}
\end{figure}

\begin{figure}
\centerline{\includegraphics[width=20mm]{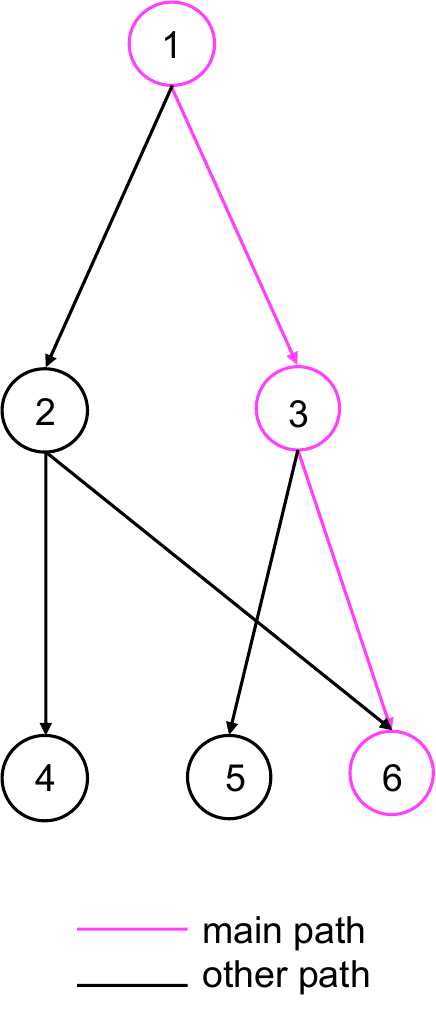}}
\caption{2nd run}
\label{main-other-path1}
\end{figure}

In this section, we randomly select paths and symbolically execute the function to infer the results of each operand in each instruction. As shown in Figure. \ref{main-other-path} and Figure. \ref{main-other-path1}, a function may contain many paths and symbolic execution might have the path explosion problem. Thus we symbolically execute the function many times. For each run, we randomly select a main path by using deep-first algorithm to symbolic execute this path. However, if we only execute one main path each run, we still may miss many instructions after many times of run (e.g., we may miss instructions 3, 4, 5 in Figure.\ref{main-other-path}, and instructions 2, 4, 5 in Figure.\ref{main-other-path1}). To cover all the instructions in each run, except the main path, we also select other paths to cover all the instructions. We do not propagate to the instructions we already encountered to avoid the path explosion problem in each run. 

Since we randomly select paths in each run, some instructions demonstrate different symbolic values. In fact, not each instruction can have different states. Only the instructions at the joint of some paths can hold different symbolic values (e.g., instruction 6 in Figure.\ref{main-other-path} and \ref{main-other-path1}). Other instructions can only hold one possible symbolic value (e.g., all other instructions in Figure.\ref{main-other-path} and \ref{main-other-path1}). The percentage of these possible symbolic values revealed is dependent on the times of run. Generally, the more time the run, the more chances we can reveal all the possible values.

\subsection{Key IR graph construction}
\begin{figure}
\centering
\begin{subfigure}[b]{0.3\textwidth}
\centerline{\includegraphics[width=\textwidth]{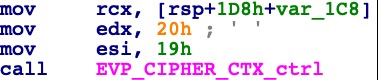}}
\caption{calling subfunction}
\label{calling subfunction}
\end{subfigure}
\hfill
\begin{subfigure}[b]{0.3\textwidth}
\centerline{\includegraphics[width=\textwidth]{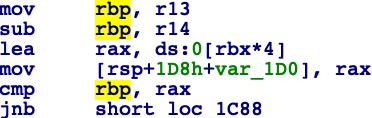}}
\caption{comparing instruction}
\label{comparing instruction}
\end{subfigure}
\hfill
\begin{subfigure}[b]{0.3\textwidth}
\centerline{\includegraphics[width=\textwidth]{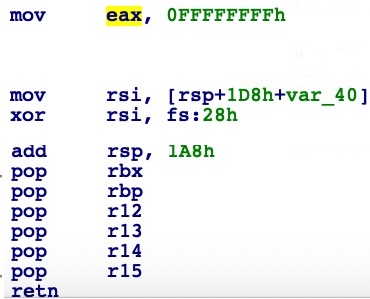}}
\caption{returning instruction}
\label{returning instruction}
\end{subfigure}
\hfill
\begin{subfigure}[b]{0.3\textwidth}
\centerline{\includegraphics[width=\textwidth]{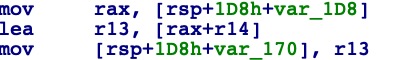}}
\caption{memory address writing}
\label{memory address writing}
\end{subfigure}
\caption{Four types of Key Instructions}
\label{four types}
\end{figure}
As we observed in the binary code, many instructions are making preparations for some other instructions, such as calling subfunction, comparing instructions, as shown in Figure. \ref{four types}. In Figure.\ref{calling subfunction}, parameters are firstly loaded into registers \textit{rcx} and \textit{edx} before calling the EVP\_CIPHER\_CTX\_ctrl subfunction. In Figure.\ref{comparing instruction}, values in register rbp are loaded from register \textit{r13} and subtracted with value in register \textit{r14}. Value in register \textit{rax} is loaded from some memory address before comparing with \textit{rbp}. In Figure.\ref{returning instruction}, value in register eax is set to 0FFFFFFFF before stack balancing instructions (i.e., pop instructions) and return instruction. In Figure.\ref{memory address writing}, value in somewhere of the memory address is loaded to register \textit{rax} and added to the value in register \textit{r14}. Then this value is written to some other memory location [rsp+1D8h+var\_170]. As we can see in all these examples, some instructions are preparing values for some later instructions. We define these preparing instructions as non-Key Instructions while the other instructions as Key Instructions. From our observation, during execution, the non-Key Instructions should propagate their values into the Key Instructions. And the Key Instructions better describe the behavior of the binary code. Similar binary code should contain similar Key Instructions with similar values (e.g., calling the same subfunction with similar parameters or comparing similar values to similar values). We define the Key Instructions to contain two types: control-flow impacting and control-flow irrelevant. Specifically, the control-flow impacting include three categories: 1. calling subfunction, 2. comparing instruction, and 3. returning instruction. We select them because Calling subfunctions leads the execution flow into other parts of the binary code. The result of comparing instructions decides which branches to take next. Returning instructions leads execution flow back to the caller function. The control-flow irrelevant instructions refer to the memory writing instructions since it does not affect the control flow.

With the result of the first module, symbolic execution, we can translate Key Instructions into Key IRs with symbolic values. Each instruction in the binary code corresponds to one Key IR node. Then we connect the Key IRs based on their control flow to form the Key IR graph. It is important to note that some nodes only contain one possible symbolic value among those Key IR nodes, while the other nodes might contain multiple symbolic values because they are the joint of multiple paths.

\subsection{Key IR graph comparison}

\begin{figure}
\centerline{\includegraphics[width=80mm]{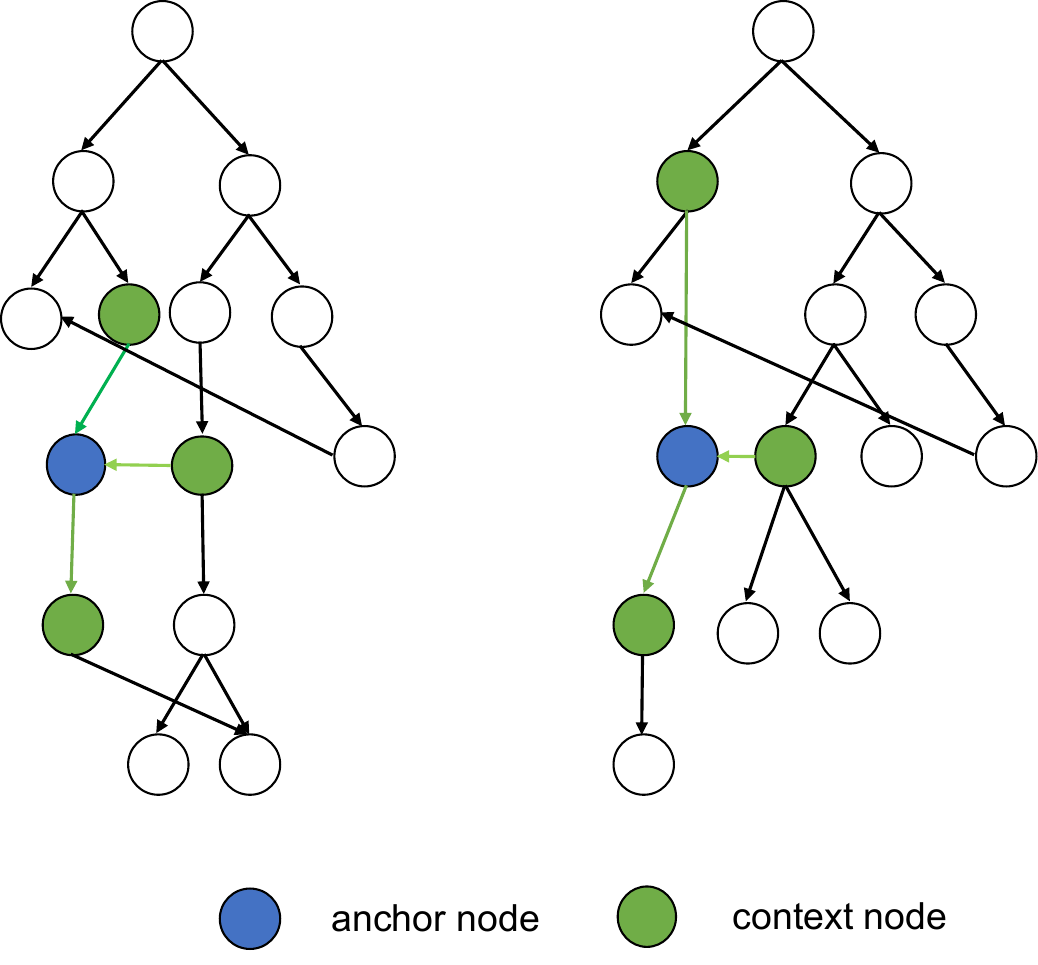}}
\caption{Key IR graph comparison}
\label{Key IR compare}
\end{figure}

Because of the impact of cross-architecture, different compiling options, and other factors, the Key IR graph might contain reordered nodes, inserted new nodes, and duplicated nodes. Sometimes Key IR graph might also lose some nodes and have other mutations because of the compilation options described in \cite{unleashing}. We propose a fuzzy method to match two Key IR graphs. To compare the similarity of two Key IR graphs, we check how many similar nodes in two graphs. To compare the similarity of a pair of node, we divide this process into two phases: 1. single node textual similarity, and 2. context similarity. In the first phase, we pick the most similar node pair from two Key IR graphs as potentially similar pairs if these two nodes' symbolic value has high textual similarity. The symbolic values have been simplified beforehand by msynth \footnote{{https://github.com/mrphrazer/msynth}}. Next, we compare the similarity of this node pairs' context. The context here refers to the neighbor nodes of a node within a given boundary. Again, we compare these neighbors' simplified symbolic value textual similarity. An example of this graph comparison is shown in Figure.\ref{Key IR compare}. Suppose we find the potentially similar node pair (blue circle) with high similarity, and the context boundary is 1. Thus we examine the neighbor nodes within this bound (green circle) and check their similarity. When detecting the context similarity, the existence of similar nodes within the context indicates the similarity. The more similar nodes are, the more similar these binaries will be. We ignore the relations between these nodes because the complexity of compilation options can significantly mutate the relations.  


\section{Implementation}
Our implementation is an IDA Pro plugin written in C++ and python. Our tool is based on IDA pro's control flow graph and its API. We have finished our implementation and on our way for evaluation.
\subsection{Symbolic execution}
\subsubsection{Symbolic values}
At the beginning of each function, we assign VAR $0$ to VAR $n$ to parameters $1$ to $N$ of this function. With the execution of each instruction, those initial tags will be represented in various expressions and are propagated to different instructions as their operands' values. The similarity of the expressions reveals similar semantics. Thus we transform binary code similarity detection into expression similarity detection.

\subsubsection{Handling loops}
During the execution, we might encounter loops. Since it is often difficult to decide the time of execution of a loop, it is challenging to propagate the values of instructions in the loop to some Key Instructions. Also, it is non-trivial to analyze which instructions' operands are invariant during loops and updated in each loop. To address these problems, we symbolically execute the instructions in a loop twice. The invariant operands keep their constant symbolic value, while the updated operands add an `ITER()' notation outside their value to highlight that their values are changed in the loop.

\subsection{Key IR graph construction}
We recover both control-flow impacting and control-flow irrelevant instructions into Key IRs. We identify those instructions if they match some patterns. The identification is implemented as a rule in the tool. For control-flow impacting instructions, we recover three types, including 1. calling subfunction instructions, 2. comparing instruction, and 3. returning instructions. For Type 1, we find the corresponding parameters for the subfunction. The patterns for matching Type 2 instruction are more complicated than other types since some compiler options can translate comparison instructions from source code to equivalent codes in an implicit way as in \cite{unleashing}. For Type 3, we find the last modification of register \textit{rax} in x86-64 or register \textit{R0} in ARM before the function returns. For control-flow irrelevant instruction, i.e., memory writing instruction, we identify them if the mnemonic is `mov' in x86-64 or `STR' in ARM, and the destination operand is a memory address.

Each Key Instruction is then transformed to a node in the Key IR graph. It is important to note that each Key Instruction might have several possible symbolic values. Thus we reserve all of them for each Key Instruction. After we recovered all the nodes, we connect them based on the control flow provided by IDA pro. The result Key IR graph has the same control flow as IDA pro's control flow. The difference is that all the non-Key Instructions are removed.

\subsection{Key IR graph comparison}
To compare the similarity of two symbolic formulas, we firstly simplify them using msynth \cite{msynth}. Msynth is originally used as a binary code deobfuscation tool. It aims to simplify very complex binary code formulas into simple formulas. We use it here to facilitate our comparison. Moreover, various compilation options can be mitigated since the resulting formula will be more similar, even the original formula may be different.

\section{Evaluation}
We have collected many open-source benchmarks used widely in existing works, i.e., OpenSSL, Coreutils, SPEC CPU2006, and SPEC CPU 2017.
We will answer two research questions in our experiment: 1. Can we effectively detect cross-architecture binary codes? 2. Can we mitigate the influence of compiling options? To answer the first question, we aim to compile our benchmark dataset on ARM and x64 architectures with different compilers such as GCC and Clang. Then we will prepare random pairs of similar and dissimilar function pairs with labels indicating their similarity. To answer the second question, we will compile benchmark dataset by using the compilation options described in \cite{unleashing} and prepare similar and dissimilar function pairs. In both experiments, the percentage of correctly detected similar and dissimilar function pairs will be regarded as the accuracy, which is to indicate the performance of our work.




\bibliographystyle{ieeetr}
\bibliography{reference}

\end{document}